\documentclass[]{spie}  

\usepackage{amsmath,amsfonts,amssymb}
\usepackage[square,sort,comma,numbers]{natbib}
\usepackage{graphicx}
\usepackage[colorlinks=true, allcolors=blue]{hyperref}

\title{The Blue Multi Unit Spectroscopic Explorer (BlueMUSE) on the VLT: End-To-End simulator `BlueSi'}

\author[a]{Martin Wendt}
\author[b,c]{Norberto Castro}
\author[c]{Sven Martens}
\author[b]{John Pharo}
\author[b]{Peter M. Weilbacher}
\author[b]{Davor Krajnović}
\author[d]{Johan Richard}

\affil[a]{Institut für Physik und Astronomie, Universität Potsdam,
          Karl Liebknecht-Str.\ 24/25, 14476 Potsdam, Germany}
\affil[b]{Leibniz-Institut für Astrophysik Potsdam (AIP),
          An der Sternwarte 16, 14482 Potsdam, Germany}
\affil[c]{Institut für Astrophysik und Geophysik, Friedrich-Hund-Platz 1,
          37077 Göttingen, Germany}

\affil[d]{Univ Lyon, Univ Lyon1, Ens de Lyon, CNRS, Centre de Recherche
          Astrophysique de Lyon UMR5574, F-69230, Saint-Genis-Laval,
          France}
\authorinfo{Send correspondence to MW via e-mail to
            \href{mailto:mwendt@astro.physik.uni-potsdam.de}{mwendt@astro.physik.uni-potsdam.de}.}

\newcommand*\farcs{\ensuremath{\overset{\prime\prime}{.}}}
\newcommand*\degr{\ensuremath{^\circ}}

\begin{document} 
\maketitle

\begin{abstract}
BlueMUSE is a blue, medium spectral resolution, panoramic 
integral-field spectrograph under development for the Very Large 
Telescope (VLT). We demonstrate and discuss an early End-To-End 
simulation software for final BlueMUSE datacube products. Early access 
to such simulations is key to a number of aspects already in the 
development stage of a new major instrument. We outline the software 
design choices, including lessons learned from the MUSE instrument in 
operation at the VLT since 2014. The current simulation software package 
is utilized to evaluate some of the technical specifications of BlueMUSE 
as well as giving assistance in the assessment of certain trade offs 
regarding instrument capabilities, e.g., spatial and spectral resolution 
and sampling. By providing simulations of the end-user product including 
realistic environmental conditions such as sky contamination and seeing, 
\texttt{BlueSi} can be used to devise and prepare the science of the instrument 
by individual research teams.

\end{abstract}

\keywords{Astronomy, BlueMUSE, Software, Simulation, data cubes, Integral-field spectroscopy, VLT, MUSE}

\section{INTRODUCTION}
\label{sec:intro}  
The integral-field spectrograph BlueMUSE will open up a new range of galactic and extragalactic science cases 
facilitated by its specific capabilities\cite{richard2019,spie_2024richard}.
In the following we demonstrate and discuss the early End-To-End simulation software \texttt{BlueSi} for final BlueMUSE datacube products.

Providing early access to simulations is crucial during the development stage 
of a new major instrument. For example, in the final preparation phase before 
the first observations with VLT/MUSE\cite{Bacon+10,2014Msngr.157...13B}, the 
consortium created a tool to generate synthetic data cubes that mirrored the 
targeted science use cases. This allowed the various science teams to test and 
refine their data analysis tools using data closely resembling the final 
observational products. Additionally, this process familiarized users with the 
general 3D IFU data format they would encounter. 
A more versatile rewrite of the simulation code was developed by MW in parallel with the 
MUSE Python Data Analysis Framework (\texttt{MPDAF}\cite{mpdaf}). 
As the framework underwent major changes, only a few science simulations were completed before the first light of MUSE. 
These included mockup cubes of sections of the Hubble Ultra Deep Field, 
as well as a comprehensive simulation of a globular cluster containing approximately 38,000 stars.
This globular cluster simulation served as a testing ground for resolving stellar populations using 3D spectroscopy in crowded field conditions\cite{kamann2013}.
The MUSE instrument was highly successful from the outset, with the 
commissioning data even leading to scientific publications. This was 
particularly evident in the study of the globular cluster NGC 6397, which 
featured the aforementioned crowded field spectroscopy\cite{husser2016} and a transverse 
science case directly motivated by earlier simulations.\cite{wendt2017} Once the real 
MUSE instrument data became available in 2014, the need for a simulation tool in its 
current state was diminished.

Future users of BlueMUSE will have access to a large pool of released tools and experience, as well as laptop-sized 
hardware capable of managing the expected data products. 
A key lesson from developing dedicated simulation tools is that they need to be available well before planning science observations,
as this is where simulations can reach their full potential. 
While software like exposure time calculators (ETC\footnote{https://eso.org/observing/etc/}) can provide estimates of expected signal-to-noise, 
the success of scientific data exploitation can only be fully tested and confirmed by deploying the full analysis chain on actual data.
To achieve this goal successfully, the simulations must generate data that closely mimics the expected output from the official data reduction pipeline. This includes accurately replicating the geometrical properties of the data cube, as well as key instrumental parameters like spatial and spectral sampling, spectral resolution, and overall cube dimensions. Additionally, the simulations must implement a range of realistic science objects, from individual point sources to crowded clusters of stars, groups of galaxies, and diffuse emission.

The End-To-End simulation software \texttt{BlueSi} provides a software interface in Python 3.x (\cite{10.5555/1593511}) to an increasing number of such
science cases as well as the implementation of the observing conditions, such as the seeing conditions and exposure times
but also the full composition of the telluric background and the resulting complex noise properties of the data.

In its current state, the simulation software package is utilized to evaluate some of the technical specifications of 
BlueMUSE as well as giving assistance in the assessment of certain trade offs 
regarding instrument capabilities, e.g., spatial and spectral resolution and 
sampling.
In the following we depict some of the first simulations that are based on actual analyses of MUSE observations to
verify the simulation's grade of realism and subsequently evaluate the benefits and differences
per science case that arise from the unique capabilities of BlueMUSE.

One of the key lessons learned from operating the MUSE instrument at the VLT 
since 2014 was the invaluable role of simulation software from the very start. 
By providing simulations of the end-user product, individual research teams 
can devise and prepare their science cases more effectively. The design of this 
software package is part of the general BlueMUSE development. The \texttt{BlueSi}
development is carried out in close cooperation with the science working groups within the BlueMUSE team.

This allows us to implement science 
objects that focus on the actual requirements and expectations regarding the 
user interface as well as the data formats.
To build and improve upon experiences with MUSE, we administered 
an organizational structure to support and guide these efforts specifically 
within the consortium. 

The fact that BlueMUSE will be a single mode instrument allows the science 
preparation teams now, as well as the scientific community later, to probe a 
comprehensive grid of observing conditions and their impact on the intended 
science for the respective object class. With typical data cube sizes of 3-6 
GiB, the data product of a simulation or a single BlueMUSE observation can well 
be handled on modern notebook- and desktop-style platforms. 
\section{End-to-End simulations}
\subsection{The BlueMUSE instrument}
BlueMUSE is a blue-optimized, medium spectral resolution, panoramic integral-field spectrograph under development for the Very Large Telescope (VLT). The instrument is currently
in Phase A development and is part of the VLT2030 instrument suite, with an expected first light at the telescope in 2031. With an optimized transmission over the wavelength rage
350-580 nm, average spectral resolution of R$\sim$3500, and a large FoV (1 arcmin$^2$), BlueMUSE will open up a new range of galactic and extragalactic science cases facilitated
by its specific capabilities. The large field-of-view of BlueMUSE is first sliced into 16 sub-fields by the field splitter, and each sub-field is further divided by 48 slices of
0\farcs3 width on the image slicer. Each slice is then image onto a detector, effectively sampling the field of view spatially with 0\farcs2 $\times$ 0\farcs3 rectangular spaxels and spectrally
with 0.66 Angstrom pixels.\cite{spie_2024richard}

\subsection{Data cubes}
A typical observation of a single pointing with the BlueMUSE spectrograph will yields datacubes with sizes of 3 GiB.
The total size breaks down as $320 \times 320 \text{ (spatial dimension)} \times 
3800 \text{ (spectral bins)} \times 4 \text{ (32bit float)} \times 2 \text{ (data and variance)}$ and is comparable to
the data generated by the current MUSE pipeline. These numbers take into account 
the common observational strategy of rotational dithering with 2-4 exposures rotated by 90 degrees per pointing.
The data cubes generated by the MUSE and future BlueMUSE data reduction pipeline as well as the data produced by
the \texttt{BlueSi} simulation are stored in the Flexible Image Transport System 
(FITS\cite{wells1981}) as a standardized and open file format.
\subsection{The BlueSi Simulation code}
The End-to-End {\bf Blue}Muse {\bf Si}mulation software \texttt{BlueSi} is written from scratch in Python 3.x
and utilizes some functionality from the by now fully established MPDAF (\cite{mpdaf}).
\texttt{BlueSi} attempts to rely on external packages only when they are fully accepted and customary and the added functionality
is deemed mandatory. This includes packages like \texttt{NumPy}\cite{numpy}, \texttt{Astropy}\cite{Astropy}, and \texttt{SciPy}\cite{2020SciPy-NMeth},
which are dependencies of the \texttt{MPDAF} package as well.

\texttt{BlueSi} produces BlueMUSE-like data products but also MUSE-like data products, however the latter is mostly
intended for the validation of the tool itself and to ease some comparisons between the data of the two instruments.

To remain accessible to a broader science community in the future, a main aspect in the design of the software is a
well defined interface to the user and a close collaboration for the integration of several
science simulations within the consortium during development.
Part of this user-friendliness is to separate the simulation core from the
parts that require input and adjustments.

\begin{figure}[h!]
\begin{center}
\includegraphics[width=0.7\textwidth]{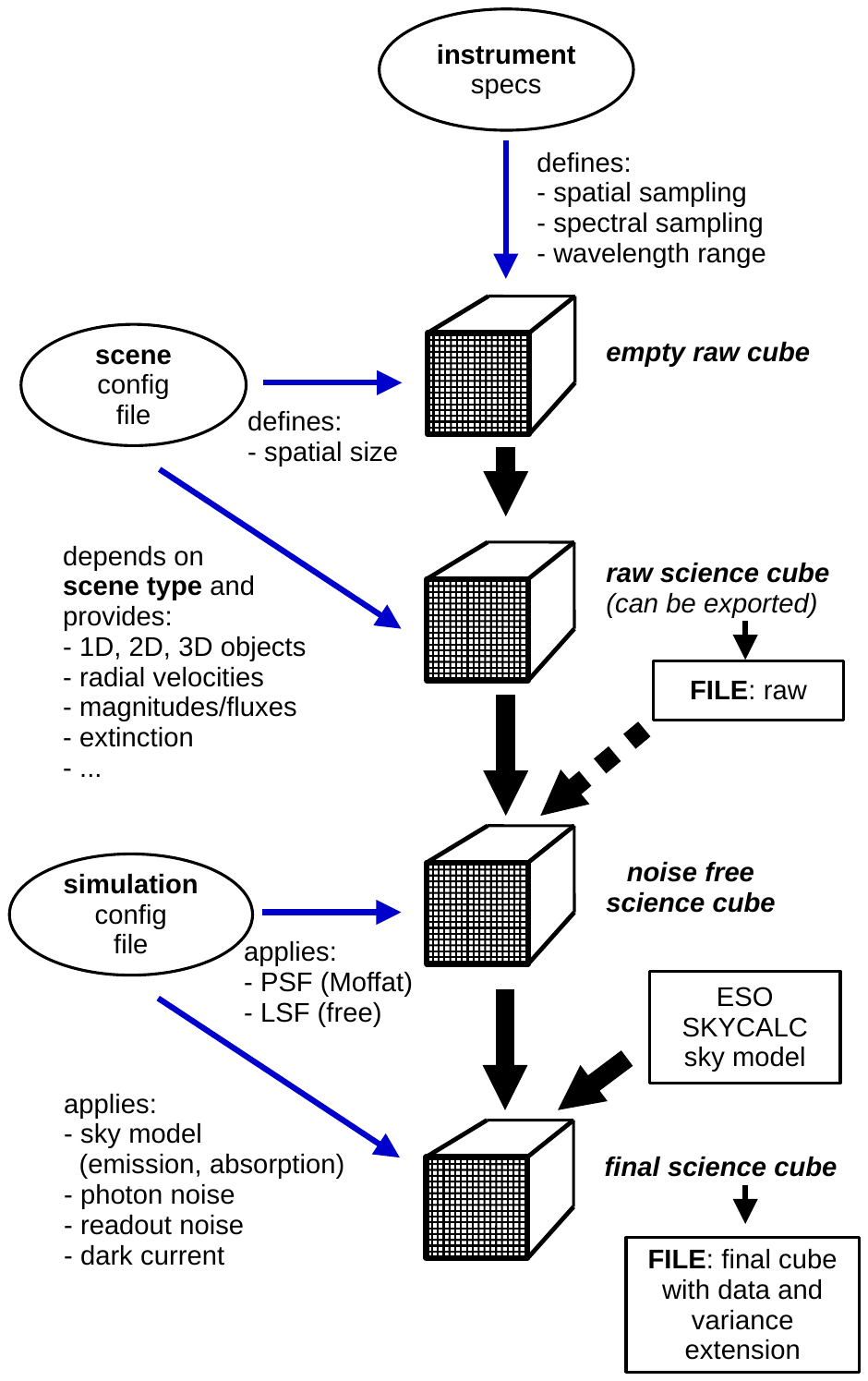}
\end{center}
\caption{Schematic of the \texttt{BlueSi} work flow.}
\label{fig_bluesi}
\end{figure}

Figure \ref{fig_bluesi} illustrates the work flow to setup and run a simulation for \texttt{BlueSi}.
The two main extern configuration files are the {\it scene config} and {\it simulation config} files.
Both are plain ASCII lists in human readable format.
The {\it scene config} file (see Appendix \ref{sec_scene_config}) contains the information
specific to the science content of the simulated cube.
Here, the user defines the type of scene, a field of view (FOV), the main input catalog as well as numerous optional 
parameters. The latter include flags to apply radial velocities, extinction correction or vacuum-to-air correction.
The main input catalog is a list of objects that describe the full science setup.
In the case of the globular cluster (see section \ref{sec_ngc3201}) that was a table with one entry per star with 
the identification number and data on positions in the sky, F606W magnitudes, radial velocities, and the
filename of the corresponding synthetic 1D input spectrum, with a total of $\sim$ 6,500 objects for this pointing.
The {\it scene config} can also contain additional information on the simulated object as well as references to
its source material and the authors of the scene.

The {\it simulation config} file  provides all the details
related to the instrument and the simulated observation and is independent on the scene setup described above.
Here all technical specifications of the simulated instrument (e.g., BlueMUSE or MUSE) 
are listed (see Appendix \ref{sec_simulation_config}).
In addition, the environmental conditions (seeing, airmass) and the observation setup (number of exposures and
exposure time per exposure) are specified.
The telluric emission and absorption is referenced from the full Cerro Paranal Advanced Sky Model
Version 2.0.9 (\cite{noll2012,jones2013}), and can read in as an external file provided from either the Python tool
or directly from the online Sky Model web interface\footnote{https://www.eso.org/observing/etc/bin/gen/form?INS.MODE=swspectr+INS.NAME=SKYCALC}.
The simulation file also assigns the relevant {\it scene config} to the simulation.

Based on the instrument specifications in the {\it simulation config} and the positional data in the {\it scene config}
file, \texttt{BlueSi} sets up a raw cube with the proper dimensions (spatial size required for the scene as well as the 
defined wavelength range) and header information (like the 
FITS World Coordinate System (WCS\cite{wcs}) based on the science scene.
The scene type specified in the {\it scene config} selects the external Python routines meant to handle the input table.
This modular approach allows for clean additions of future science object types that potentially require customized
processing. For each type there is a dedicated routine to prepare the raw data cube (i.e., deriving the overall spatial extent of a described scene) as
well as a routine to arrange and add the data of the scene to the raw cube.
In the depicted case of the globular cluster scene, all input files are 1D spectra, for which a small 3D cube is
created to handle their proper positioning at sub pixel precision. In this step radial velocity shifts, magnitude flux
scaling, vacuum-to-air transformation, extinction correction and resampling to the cube grid are carried out.
Depending on the requirements, this step is the most time intensive (up to a few hours for the full globular cluster scene) and the 
intermediate cube can be exported at this stage.

Next the observational conditions are applied in the simulation. The cube is convolved image layer by layer with a PSF 
described with the wavelength dependent Moffat function.
Here, the impact of rectangular pixels can be simulated.
To implement the LSF, the cube is convolved with a freely defined function
(in this case a plain Gaussian) in spectral direction. 
Now, the sky model data is converted to the simulated cube format and added to the flux levels before the photon noise per data point
is computed, which depends on the object count, the sky count and the dark count. 
Depending on the selection in the {\it simulation config} the sky flux itself is either removed or remains in the data 
extension of the cube. The readout noise is also applied according to the specifications, taking into account saturation 
effects. All noise related aspects depend on the particular arrangement of exposures time and number of exposures.
The selection of the simulated instrument in the {\it simulation config} affects the total throughput, that is also being simulated.
The true variance of the data is optionally stored as an extra extension in the data cube (effectively doubling its size).
In the end, the final data cube is being stored along with a white image, and all the relevant entries from the {\it scene config}
and {\it simulation config} saved as keywords in the FITS header.

All the steps listed in the last paragraph can be applied directly to a previously generated cube to save a significant amount of time
and to quickly render multiple simulations with different observation conditions or strategies (the computations for sky, PSF, LSF, and noise levels take a few minutes for a normal scene).

In the following we present the current status of different types of scenes that are currently implemented.
The aforementioned globular cluster scene also acts as a benchmark for the simulation code itself (see section \ref{sec_ngc3201}),
a crowded scene of young hot stars and diffuse ISM gas, inspired by NGC2070 (section \ref{sec_ngc2070}) and a science scene containing
Lyman-$\alpha$ emitters at various redshifts and flux levels as an early type of 3D input objects in section \ref{sec_lae}.
\section{Simulated observations}
\subsection{Globular cluster NGC 3201}
\label{sec_ngc3201}
The  science scene we simulated first is based on a globular cluster, namely NGC 3201.
NGC 3201 was thoroughly studied in \cite{Giesers2019} and constitutes an ideal object to achieve two goals at the same time for \texttt{BlueSi}.
On the one hand, the existing analysis enables us to create essentially an exact copy via simulation to evaluate the software and on 
the other hand, it can directly outline the benefits and changes for a potential observation with the upcoming BlueMUSE instrument.
\subsubsection{Scene setup}
\begin{figure}[h!]
\begin{center}
\includegraphics[width=0.75\textwidth]{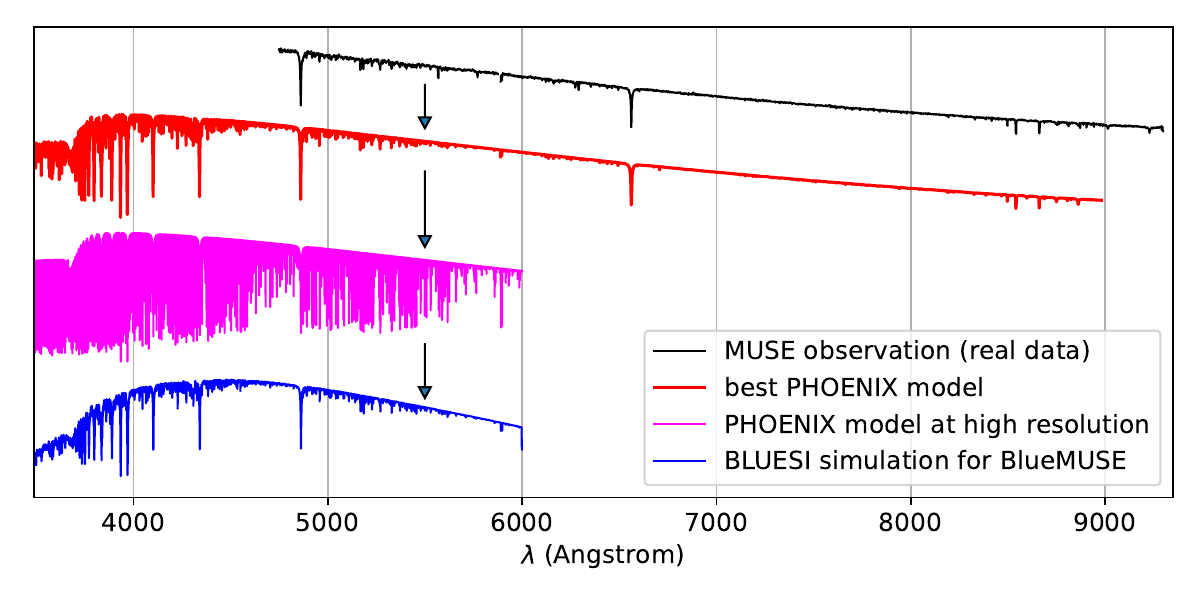}
\end{center}
\caption{Example of the procedure from original MUSE observations of stellar objects to synthetic spectra for BlueMUSE ({\it top to bottom}).}
\label{fig_gcscene}
\end{figure}

The catalog that makes up the \texttt{BlueSi} scene of the globular cluster NGC~3201 is based on a MUSE field of view of that cluster. 
Figure \ref{fig_gcscene} illustrates the whole process. The stars in the crowded field of the globular cluster were initially simultaneously de-blended and extracted 
via the \texttt{PampelMUSE} Software (\cite{kamann2013}).
The stellar parameters of the included stars are based on averages of results from MUSE observations of these stars by 
fitting against the Göttingen Spectral Library (\cite{husser2013}) that consists of a range of spectra based on stellar atmosphere models.
On average 20, with up to 60, individual observations contribute to the best match library model for each individual star.
The analysis also yielded the radial velocities per object which are particularly high for this globular cluster due to the high relative velocity of the cluster with respect to the solar system (around 490 km/s).
For the \texttt{BlueSi} simulations, the stellar spectra are then generated by interpolating the Göttingen Spectral Library using the parameters effective temperature, surface gravity and metallicity which were derived from the numerous science observations.
Conveniently, the synthetic spectra are available at a much higher resolution (R $\sim$ 500,000) and down to the 3,000 \AA{} in wavelength 
and thus extent even below the blue range of BlueMUSE.
A total of $\sim 6,500$ synthetic spectra contributed to the full simulation of NGC 3201. The comparison with actual MUSE observations was later restricted
to one of the pointings to be less affected by the mixture of different observation conditions for the pointings (most notably, the sky background and seeing).
The analysis tool chain for the globular cluster applies its own sky reduction which is why the atmospheric emission and absorption features were left in the
simulated data cube (sky removal is usually being dealt with in the data reduction software, and therefore, the sky is also not present in the simulated data cubes).
The simulated sky was modeled based on the environmental conditions of the original MUSE observations of the simulated region.
For the simulation of the MUSE data, a wavelength range of 4750-9300Å with a sampling of 1.25Å and a resolution of  $R=1800$ (at 5000Å) was used.
The PSF was modeled as a Moffat function corresponding to a FWHM of 0\farcs63" at 500 nm.
Identical to the real observation, three exposures of 200 seconds (600 sec in total) were simulated -- taking into account a readout noise of 3 electrons and a dark level of 3 electrons per hour.
In addition a global extinction of E(B-V) = 0.19 for the whole scene was applied.
To match the data from the MUSE data reduction system, the final data cube wavelength calibration was provided in air, applying the required atmospheric
dispersion as in \cite{edlen1966}.

\begin{figure}[h]
\begin{center}
\includegraphics[width=0.45\textwidth]{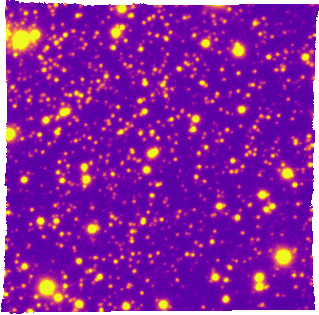}
\includegraphics[width=0.45\textwidth]{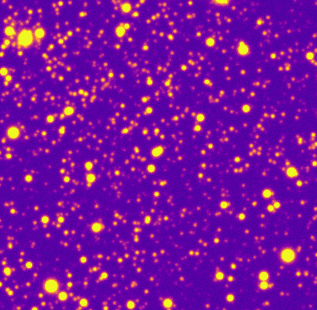}
\end{center}
\caption{{\it Left:} Original VLT/MUSE observation of NGC 3201, {\it right:} Matched \texttt{BlueSi} simulation of MUSE data.}
\label{fig_bluesi3201}
\end{figure}

Figure \ref{fig_bluesi3201} shows a single layer of the simulated cube in comparison to the corresponding original observation.
The level to which both match reflects the extent of the original scientific analysis and the ability to de-blend such crowded fields as well as the
degree of realism that \texttt{BlueSi} reached with this simulation.
The main reason for differences at this stage are unresolved sources and sources that we were unable to analyze in the original data.

For the simulations in BlueMUSE mode, the same parameters regarding sky and exposure were applied.
To match the BlueMUSE specifications, the wavelength range was set to 3500.0-6000Å with a sampling of 0.66Å and a resolution of  $R=4000$ (at 5000Å).

\subsubsection{Extraction and fit}
To extract spectra of single stars from a simulated datacube, we use \texttt{PampelMUSE} (\cite{kamann2013}), mimicking the original analysis of
the MUSE observations of NGC 3201.
Throughout the analysis the same settings as for real MUSE observations as described in \cite{kamann2016} and \cite{husser2016} were applied.

To derive the stellar parameters from the extracted spectra, we employ \texttt{spexxy} which implements a full spectral fitting approach, described in detail in \cite{husser2016}. The stellar spectra are fitted against the Göttingen Spectral Library. For simulated MUSE spectra, we use a grid of models that has been convolved with an empirical LSF of the MUSE instrument. To analyze the BlueMUSE spectra, the grid of models is convolved with a Gaussian with a FWHM of $1.14\,$\AA, which corresponds to the anticipated spectral resolution $R\sim 4000$ of BlueMUSE \cite{richard2019}.

\begin{figure}[h!]
\begin{center}
\includegraphics[width=0.3\textwidth]{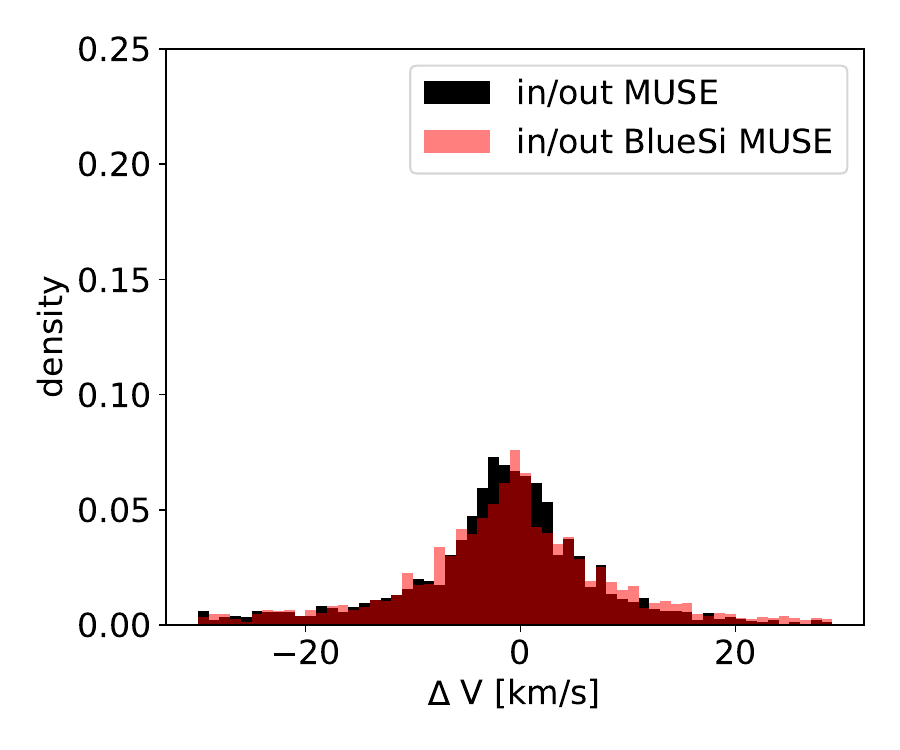}
\includegraphics[width=0.3\textwidth]{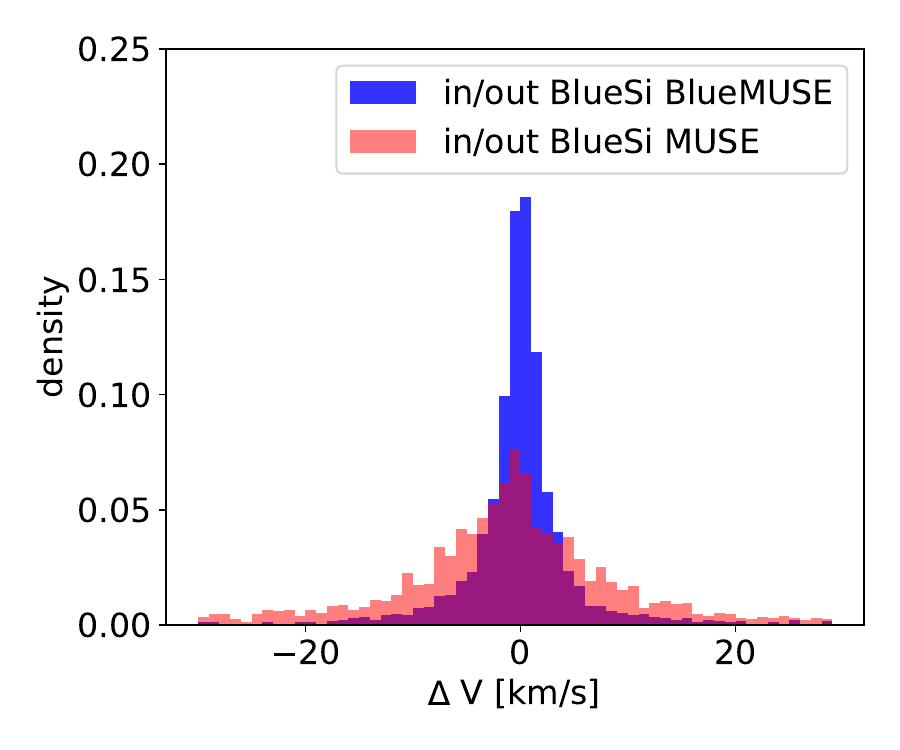}
\includegraphics[width=0.3\textwidth]{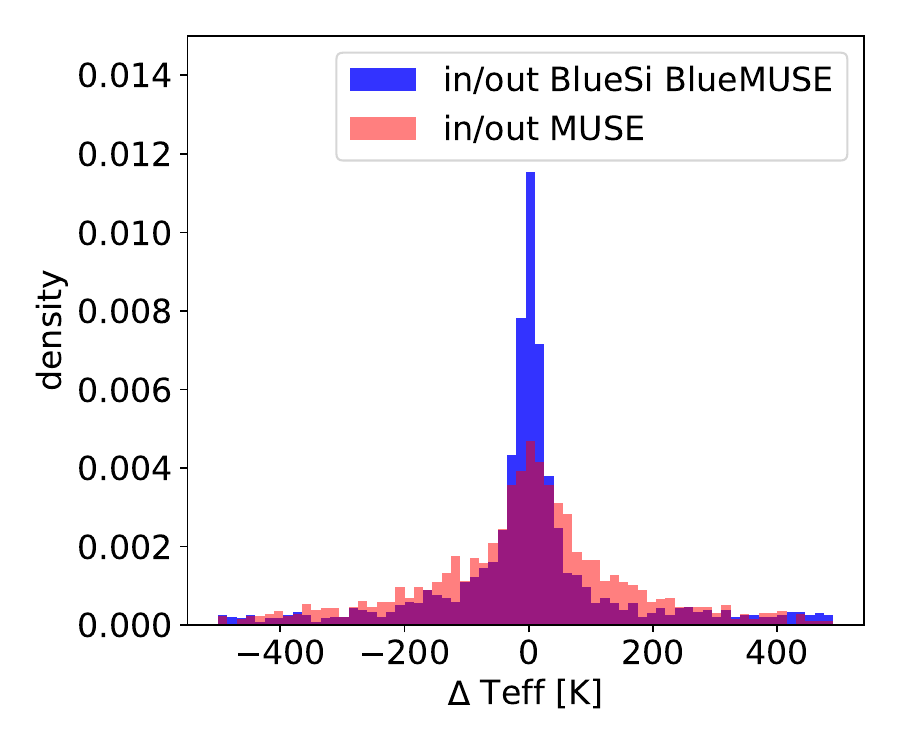}
\end{center}
\caption{Globular cluster scene: The velocity offsets between the model velocities in the \texttt{BlueSi} simulations and the radial velocities recovered
from the data cube analysis. {\it Left:} from a set of only three MUSE observations and the corresponding \texttt{BlueSi} simulation. 
{\it Centre:} in comparison to the same analysis on a \texttt{BlueSi} simulation for the BlueMUSE instrument.
{\it Right:} deviations of the fitted effective Temperature T$_\text{eff}$ to the input.
}
\label{fig_gc_vel}
\end{figure}

Figure \ref{fig_gc_vel} summarizes some aspects of the analysis.
The direct comparison of the individual derived radial velocities between the original MUSE observations and the corresponding \texttt{BlueSi} realization thereof shows
the general success in that regard. For both, the precision with which velocities could be recovered is a mere fraction of the pixel size (MUSE ~ 75 km/s per pixel at 5000\AA).
It is noteworthy, that the input parameter against which we compare both data cubes were to some extent derived from this particular MUSE observation
itself. This bias is likely balanced by the non-perfect wavelength calibration of the physical instrument whose implications are not integrated in the \texttt{BlueSi} software as of now.
As expected, the benefit of BlueMUSE's higher resolution and sampling by a factor of $\approx 2$ shows in the success of recovering the input velocities.
The aspect of wavelength calibration precision will play a more pronounced role in the blue wavelength range and is part of the future improvements intended for \texttt{BlueSi}.
The extended wavelength range in the blue will yield additional diagnostic power besides the gain in resolution, given that there are plenty of stellar features
in the additional part of the spectrum (see, e.g., Figure \ref{fig_30dor_bluemuse_fit}).
The \texttt{BlueSi} simulations of this cluster with the current specifications of the BlueMUSE instrument, demonstrate it's ability to derive the effective temperatures
of the stars at a significantly higher precision as well (see Fig.\ref{fig_gc_vel}, {\it right}).

\subsection{The most massive stars in NGC2070}
\label{sec_ngc2070}
BlueMUSE will open a new and unique gateway to studying blue massive stars. It will enhance and build upon the capabilities of MUSE in terms of multiplexity, sensitivity, large field-of-view, and spectral resolution, specifically exploring the 3500-5800 Å optical range. This spectral range will provide access to a large set of chemical transitions (e.g., oxygen, carbon, and nitrogen) of key elements essential for constraining stellar evolution. Massive stars are highlighted as a key science case in the BlueMUSE White Paper and represent one of the main scientific drivers for this new instrument at the VLT. MUSE has already demonstrated the remarkable potential of state-of-the-art integral-field spectrographs for de-blending and analyzing crowded fields of massive stars. However, the wavelength range explored by BlueMUSE will be revolutionary for the characterization of blue massive stars.

\texttt{BlueSi} simulations of densely populated massive star clusters are crucial to test BlueMUSE's ability to resolve dense stellar fields and provide accurate stellar parameters and chemical abundances. The massive core of 30 Doradus (also known as NGC 2070) serves as an ideal test-bench for these simulations, offering a challenging environment to demonstrate the instrument's capabilities.

\subsubsection{Scene setup}

The initial \texttt{BlueSi} NGC2070 catalogue was built using the 518 stars extracted from the Science Verification cubes obtained in 2014, mapping the core of 30Dor with a mosaic of four points and four exposures of 600 seconds each\cite{Castro2018}. The stars were analyzed using the stellar atmosphere code FASTWIND\cite{Puls2005}. Only those stars with S/N $\ge$ 50 (~300) were published\cite{Castro2021}; nonetheless, to reach a more realistic view of a massive stellar cluster, the 518 stars with an estimation of effective temperature and gravity were used in the initial \texttt{BlueSi} 30Dor catalog. Additionally, we applied individual radial velocities for each target around the systemic velocity of NGC2070, 265km/s\cite{Castro2018}.

Each star in the catalog was associated with a synthetic spectral energy distribution template based on the effective temperature and gravity. We used Tlusty models\cite{Lanz2007} for stars hotter than 10000K and Phoenix libraries\cite{husser2013} for any stars cooler than this temperature. Tlusty and Phoenix libraries include a complete set of chemical transitions based on temperature and gravity, essential to carry out the simulations.

Although the initial catalog demonstrated the effectiveness of \texttt{BlueSi} in simulating a massive cluster,  518 stars are not enough to test a highly dense region such as NGC2070. To address this, we increased the stellar population using the HST photometric catalog from the Hubble Tarantula Treasury Project\cite{Sabbi2016} (HTTP), which provided almost 3000 candidates within the region covered by the Science Verification 2014 MUSE data.

We cannot accurately simulate massive stars in NGC2070 based on photometry alone, which is why we opted to create a synthetic cluster inspired by 30Dor, which would offer a suitable environment for testing stellar extraction techniques and analysis tools in a dense setting. To assign effective temperature and gravity values to each star, we simulated the isochrones of a massive cluster using the SYCLIST tool\footnote{https://www.unige.ch/sciences/astro/evolution/en/database/syclist/}. Utilizing synthetic colors generated by SYCLIST at the distance of the LMC (49 kpc\cite{Pietrzynski2013}) alongside the observed HTTP photometry, we derived temperatures and gravities for our targets. Subsequently, Tlusty and Phoenix models were associated with each target, following the same approach taken in the first catalog.

\subsubsection{Simulation}
For the \texttt{BlueSi} runs, an observation consisting of 4 $\times$ 600 = 2,400 seconds was simulated.
The identical exposure times and Cerro Paranal Advanced Sky Model were chosen for the simulation of data from the MUSE and future BlueMUSE
instrument. The technical specifications of both were identical to those described in section \ref{sec_ngc3201}.
Additionally, the scene was simulated at a projected distance of 75 kpc to test the instrument's expected resolving power as well as the 
ability of the analysis framework to de-blend and independently analyze the objects in the crowded field (see section \ref{sec_distance}).

\begin{figure}[h!]
\begin{center}
\includegraphics[width=0.8\textwidth]{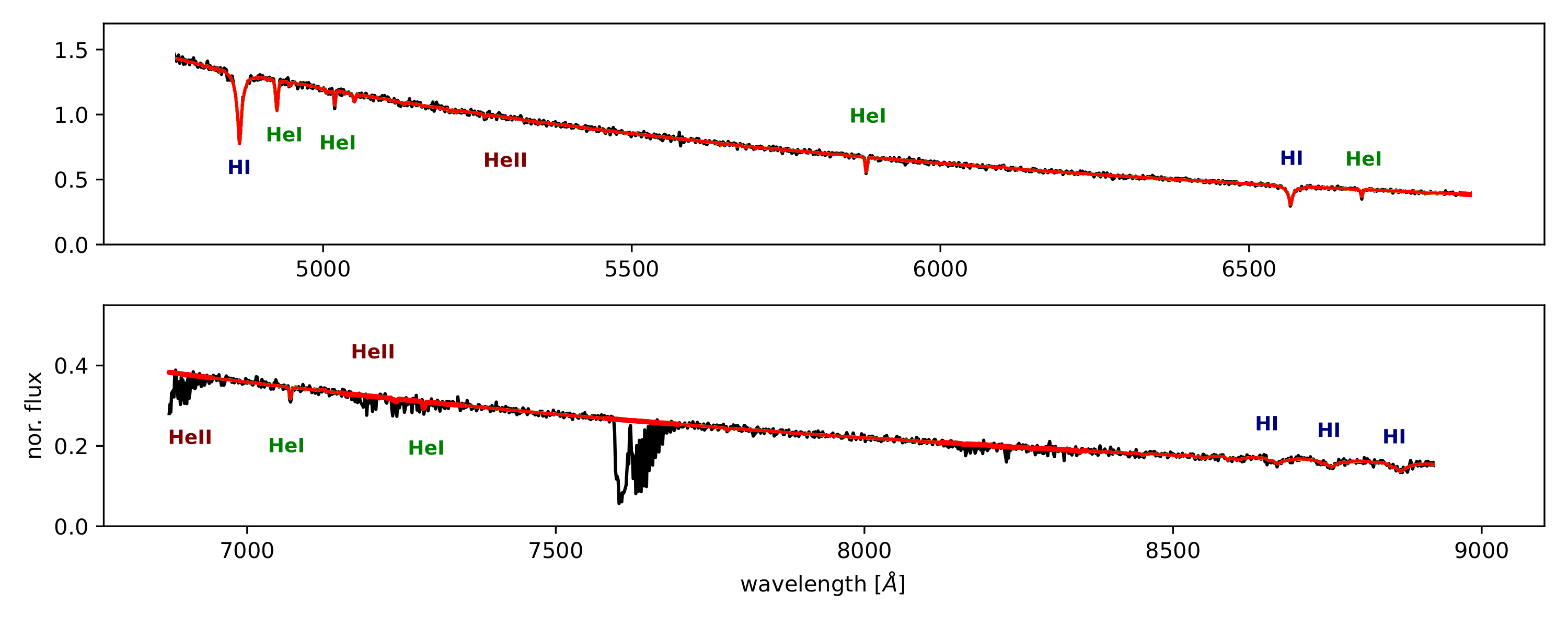}
\end{center}
\caption{Extracted spectrum of a single star ({\it black}) from the \texttt{BlueSi} simulation of a MUSE observation with the best template fit ({\it red}).
Absorption features of prominent stellar transitions in the covered wavelength range are labeled. The simulated data includes the strong
telluric A band absorption.}
\label{fig_30dor_muse_fit}
\end{figure}
\begin{figure}[h!]
\begin{center}
\includegraphics[width=0.8\textwidth]{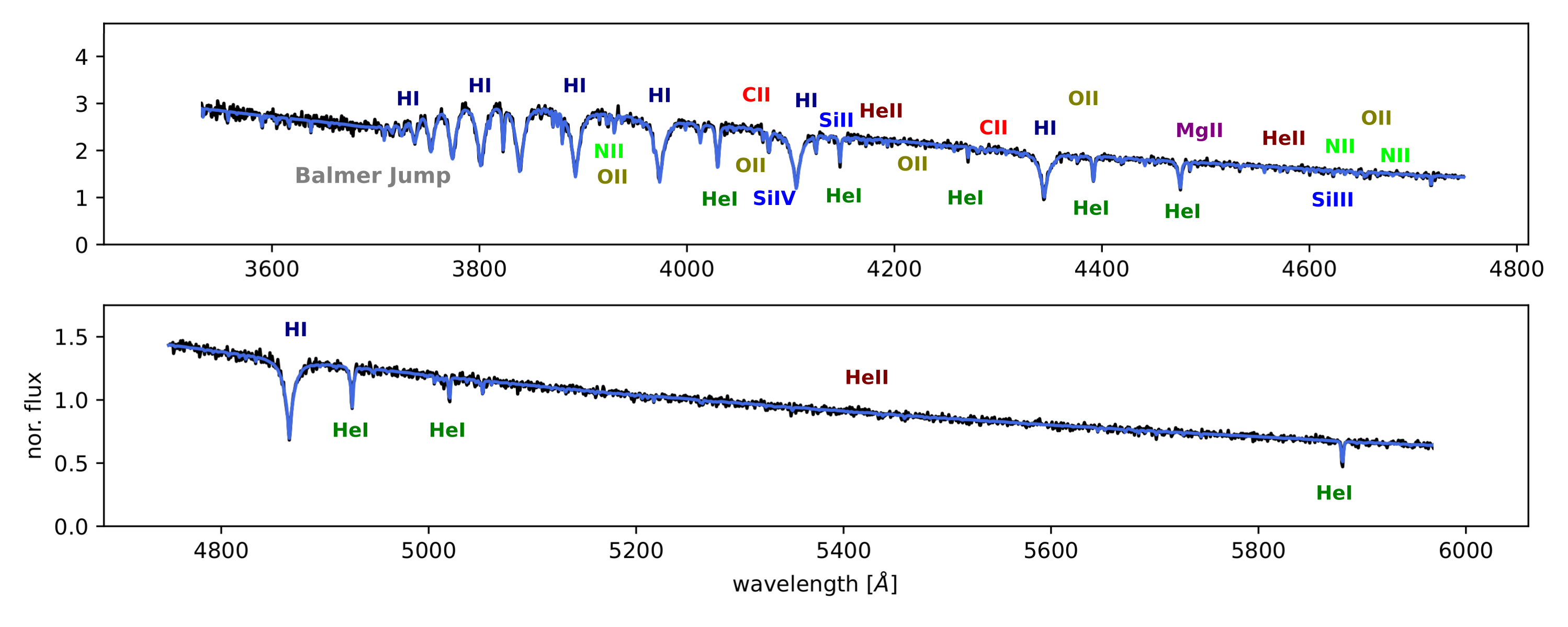}
\end{center}
\caption{Extracted spectrum of a single star ({\it black}) from the \texttt{BlueSi} simulation of a future BlueMUSE observation with the best template fit ({\it blue}).
Absorption features of prominent transitions in the covered wavelength range are labeled. There is no strong
telluric absorption in this wavelength region.}
\label{fig_30dor_bluemuse_fit}
\end{figure}
\subsubsection{Analysis}
\label{Sect:analysis}
The stellar content in both \texttt{BlueSi} datacubes was extracted using the \texttt{PampelMUSE} code. \texttt{PampelMUSE} successfully performed the extraction in both MUSE and BlueMUSE simulated data cubes, recovering approximately 90\% of the input sources. Figures~\ref{fig_30dor_muse_fit} and \ref{fig_30dor_bluemuse_fit} display examples of two simulated stars extracted from the MUSE and BlueMUSE cubes, respectively.

The extracted spectra were subsequently analyzed using the ULYSS code\cite{Koleva2009}, employing the same Tlusty+Phoenix grids utilized in the input catalog. In this study, ULYSS generates a linear combination of models that best fits the input data\cite{Roth2018}. Effective temperature, gravity, and errors were estimated from the average and standard deviations of the optimal models, weighted according to ULYSS output.

\begin{figure}[h!]
\begin{center}
\includegraphics[width=0.75\textwidth]{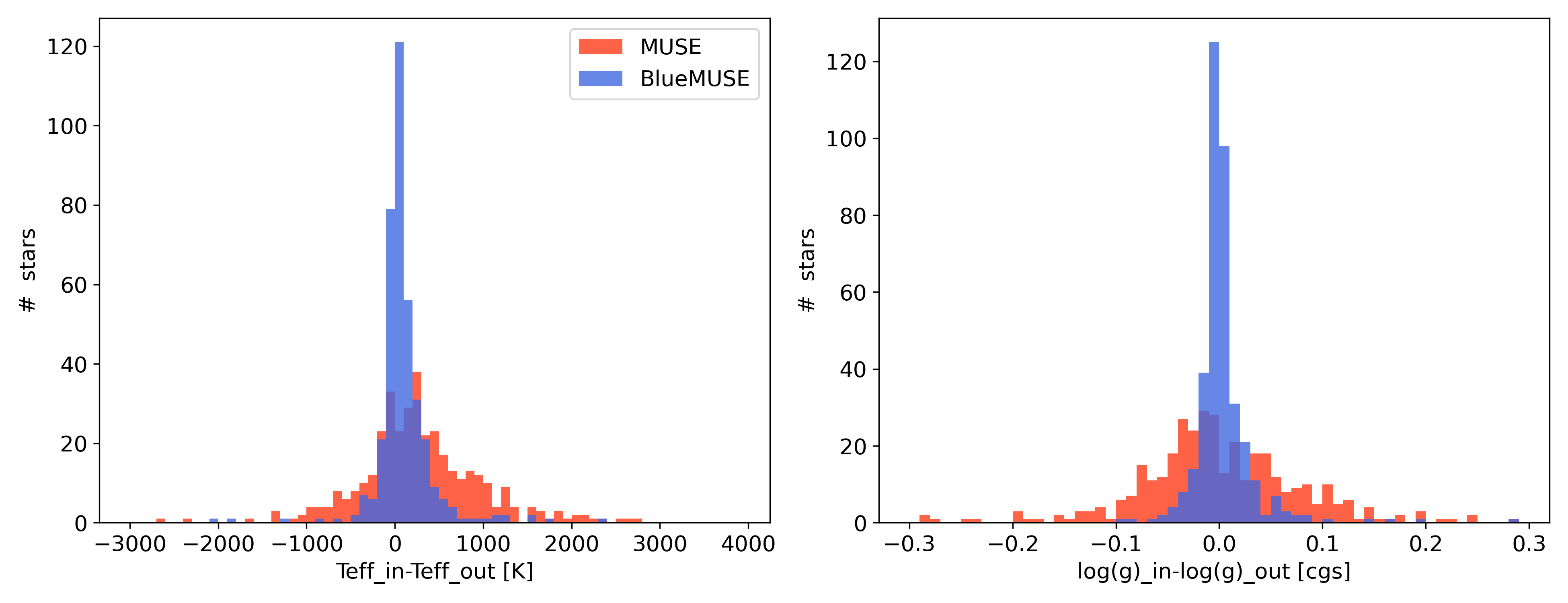}
\end{center}
\caption{Comparison of the input parameters of the simulation to the result of the analysis of the synthetic data cube for the effective Temperature T$_{\text{eff}}$ and log(g) for simulations of MUSE observations ({\it red}) and BlueMUSE ({\it blue}).}
\label{fig_logg}
\end{figure}

Figure~\ref{fig_logg} shows the difference between the input temperature and gravity and the values recovered by the ULYSS code on the left and right panels, respectively. These preliminary results indicate a significant improvement in the analysis of BlueMUSE data compared to the MUSE simulation, evidenced by a less scattered distribution and smaller errors. As it was mentioned, the wavelength range provided by BlueMUSE is more suitable for analyzing blue massive stars than that of MUSE. This is also illustrated in Figures~\ref{fig_30dor_muse_fit} and \ref{fig_30dor_bluemuse_fit}, where the wavelength range covered by BlueMUSE includes a larger number of chemical transitions used as anchors to determine stellar parameters and chemical compositions.
\subsubsection{Projection of crowded fields at other distances}
\label{sec_distance}
Massive stars are the primary chemical and dynamical engines of galaxies. Studying these stars beyond the Magellanic Clouds is crucial for understanding the role of low
metallicity environments in their formation and evolution, serving as a proxy for the low metallicity conditions of the early Universe\cite{Garcia2021}. \texttt{BlueSi} can be used to
simulate these massive clusters at different distances and test the limits of the instrument and analysis tools.

\begin{figure}[h!]
\begin{center}
\includegraphics[width=0.4\textwidth]{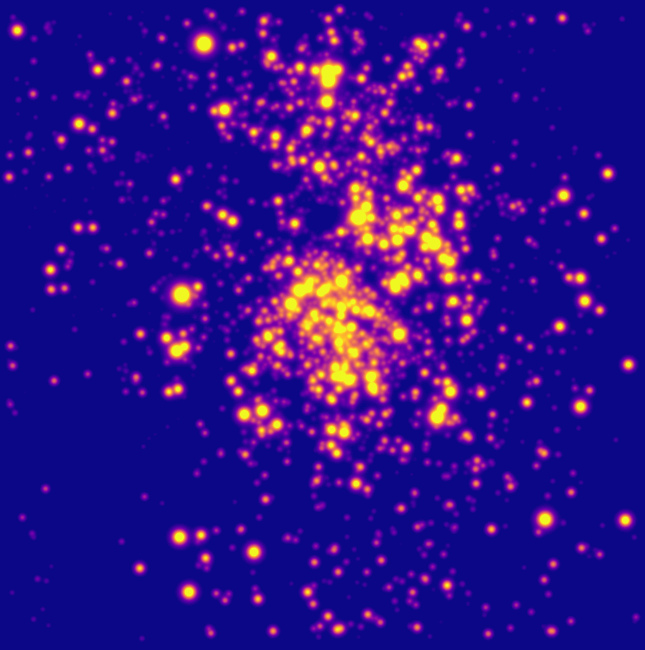}
\includegraphics[width=0.4\textwidth]{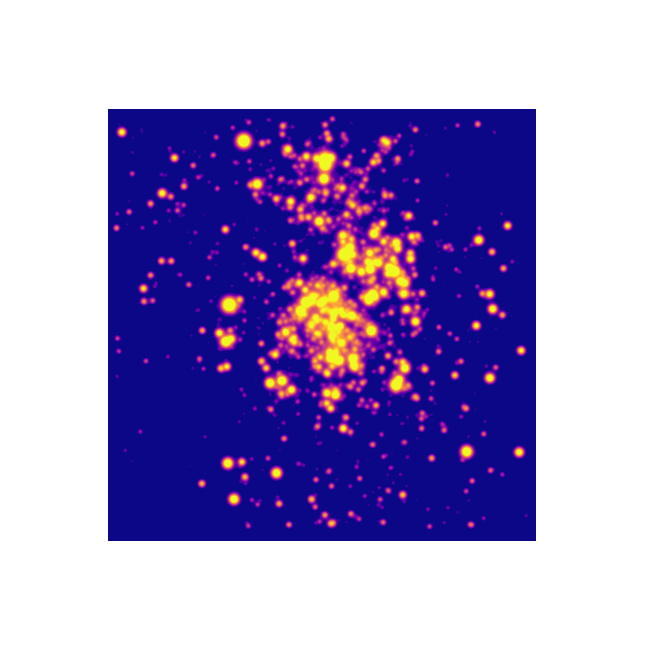}
\end{center}
\caption{White images of the 30 Doradus inspired object simulated at different distances. {\it Left:} natural, spanning about 30 parsec at the distance of 49 kpc, {\it right:} projected to a distance of 75 kpc. The logarithmic flux scale of these images conceals the impact on the visual magnitudes.}
\label{fig_30dordistance}
\end{figure}

We performed a simulation of our mock NGC2070 cluster, placing it at 75 kpc (65\% farther away) (Fig.~\ref{fig_30dordistance}). The \texttt{PampelMUSE} code was still able to recover 72\% of the input
targets. The analysis of these targets is ongoing.
\subsection{Expected Detectability of Faint Lyman-$\alpha$ Emitters}
\label{sec_lae}
Detection of a large population of faint Lyman-$\alpha$ emitting galaxies (LAEs) will be a key extragalactic science use case for BlueMUSE\cite{richard2019}. The Lyman-$\alpha$ emission line in hydrogen is a common potential observational tracer for recent star formation \cite{sobral2019}, AGN activity, and structure and dynamics of the circumgalactic medium (CGM) in galaxies (e.g., \cite{galbiati2023, lofthouse2023}). Properly constraining the population of LAEs in different epochs of the universe can therefore provide critical information on the evolving characteristics of galaxies and their CGM. Observations with BlueMUSE will expand deep coverage of LAEs from $z>3$ with MUSE toward the epoch of cosmic noon ($1<z<3$), where the cosmic rate of star formation in galaxies reached its peak\cite{madau2014}. It is also expected that the number of LAEs detectable with BlueMUSE will be significantly higher than in comparable MUSE surveys, given the reduced surface brightness dimming at lower redshift. 

Lyman-$\alpha$ is a resonantly scattered emission line\cite{partridge1967}, which in practice means it is typically observed not as a compact source but in spatially extended halos with broad, complex spectral profiles (e.g., \cite{verhamme2006,matsuda2012,momose2014,leclercq2017,claeyssens2019,blaizot2023}). This contains physical information, since the observed halos will depend on the specific configuration of Ly$\alpha$ photons scattering through the host galaxy's CGM, but it also has important implications for the observation of LAEs. For a given intrinsic Ly$\alpha$ line luminosity, the spatial and spectral distribution of the emission can strongly affect the detectability of the Ly$\alpha$ halo (LAH) in a given survey\cite{herenz2019}. Understanding the relationship between LAH shapes and LAH detectability will be key to making accurate predictions for the cosmic noon LAH populations BlueMUSE will observe. \texttt{BlueSi} provides the means to begin testing this relationship.

\subsubsection{Scene setup}

As a simple test case of \texttt{BlueSi} for this application, we inserted a small grid of LAH models into a \texttt{BlueSi} scene. MUSE observations of LAHs at high $z$ have demonstrated that the LAH spatial profile can be reasonably modeled in spatially-resolved observations with two exponential disk components: one representing a more compact, continuum-like Ly$\alpha$ emission, and a second representing the extended halo\cite{wisotzki2016}. The spectral profile can be modeled with a skewed Gaussian function\cite{shibuya2014,bacon2023}. 

\begin{figure}[h]
    \centering
        \includegraphics[width=0.4\textwidth]{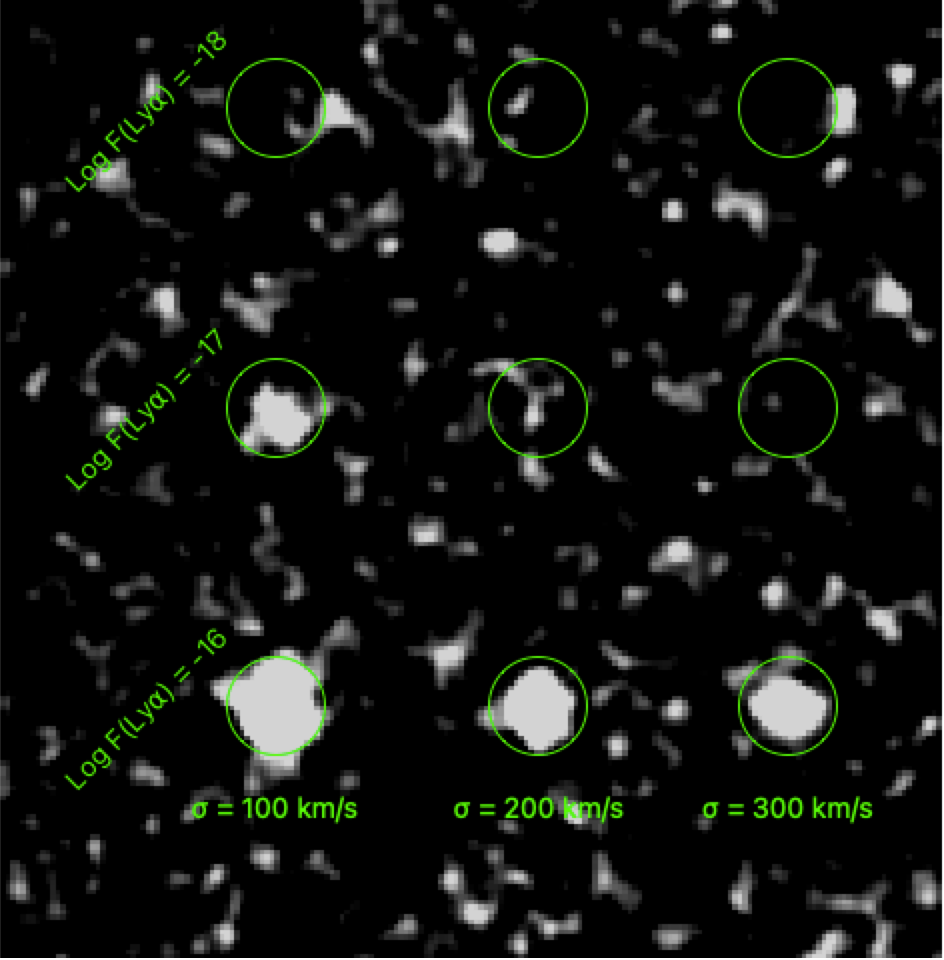} \\
        \includegraphics[width=0.7\textwidth]{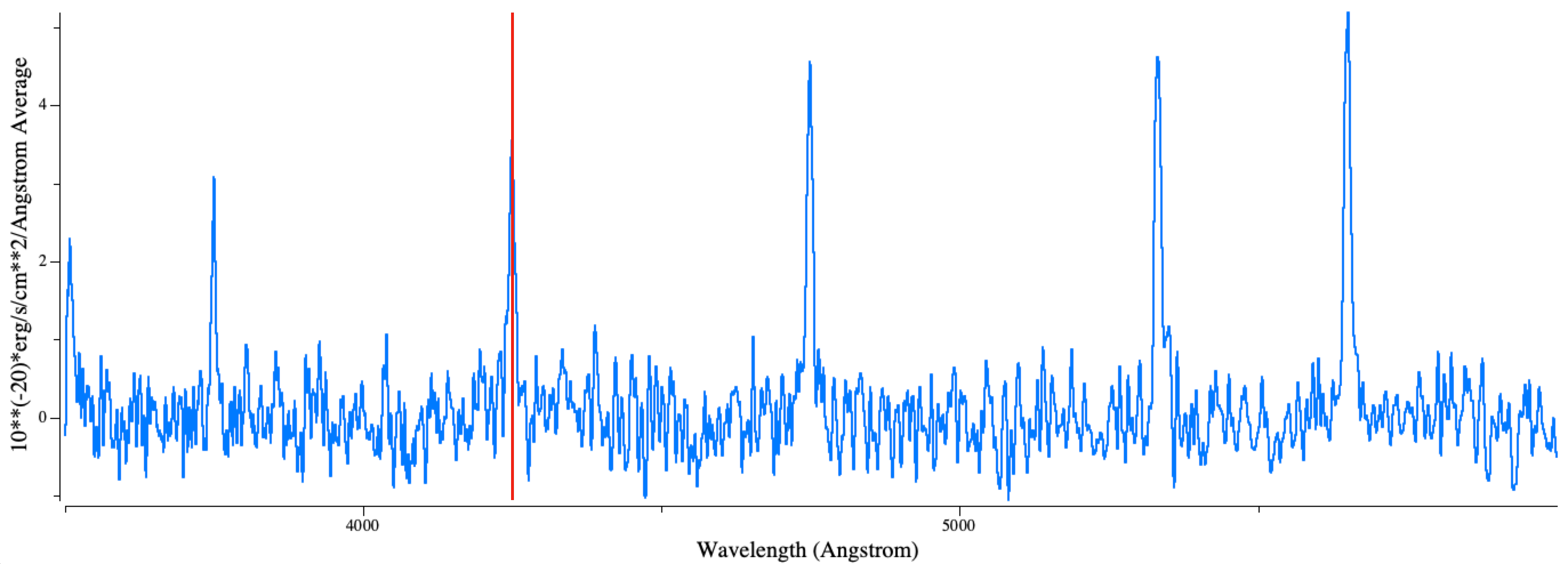} \\
    \caption{\textit{Top:} A wavelength slice of the signal-to-noise cube derived from the MUSCATEL-like \texttt{BlueSi} scene with the inserted grid of LAHs. The slice is at $\lambda=4250$ \AA\, and circles indicate the individual LAHs. Each row has a given intrinsic Ly$\alpha$ flux, and each column has a given spectral line width $\sigma$. \textit{Bottom:} Example spectrum of an LAH from the \texttt{BlueSi} simulated cube, showing the six wavelengths insertions of the Ly$\alpha$ line. The vertical red line shows the location of the wavelength slice depicted in the top panel.}
    \label{fig:lah_example}
\end{figure}

With this prescription, we developed a small test grid of models varying three parameters: the intrinsic emission line flux ($\log F_{Ly\alpha} \in [-16,-17,-18]$ in erg s$^{-1}$ cm$^{-2}$) and the Gaussian line width in the spectral profile ($\sigma \in [100,200,300]$ km/s) centered at a range of wavelengths spanning BlueMUSE spectral coverage ($\lambda_0 \in [3510,3750,4250,4750,5333,5650]$ \AA). Each model LAH is constructed first as a mini cube and then placed in the \texttt{BlueSi} scene (See Figure \ref{fig:lah_example}). The scene was constructed to imitate shallow field observations from the MUSCATEL MUSE survey\footnote{ESO program ID 1104.A-0026(B)}, with 4 exposures of 25 minutes. We passed the resulting simulated datacube to the Line Source Detection and Cataloging software (LSDCat\cite{herenz2017a,herenz2023}), a filtering tool used to recover Ly$\alpha$ emission in previous surveys with MUSE. LSDCat takes as input a flux cube and variance spectrum and produces a signal-to-noise ($S/N$) cube from which we were able to measure the expected $S/N$ recoverable for each model.
\begin{figure}[h]
    \centering
    \includegraphics[width=0.7\textwidth]{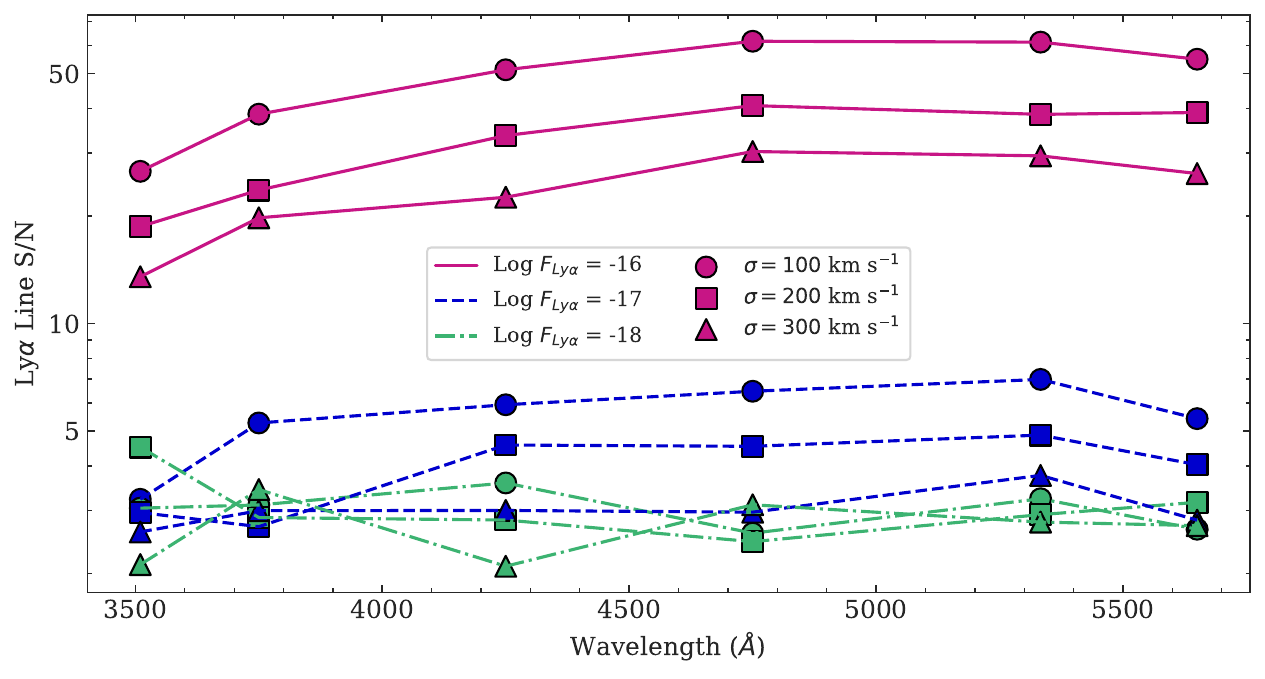}
    \caption{The recovered signal-to-noise ($S/N$) from the grid of LAHs at different wavelengths. Intrinsic line flux (-16,-17,-18) is given by color (purple, blue, green), and the line width (100,200,300 km/s) is given by different symbols (circle, square, triangle). This demonstrates both the impact that parameters such as the line width can have on the recovered $S/N$, and how \texttt{BlueSi} may be used effectively to predict the detection of different LAHs.}
    \label{fig:lah_SN}
\end{figure}
\subsubsection{Results and Future Applications}

Figure \ref{fig:lah_SN} shows some initial results from the source recovery experiment. While the main driver in the recovered $S/N$ is unsurprisingly the intrinsic line flux, the test also shows significant differences in line recovery at the same intrinsic flux but with different line widths: for a detection threshold of $S/N > 5$, a common cutoff in emission line studies, only the narrowest line profile at $\log F_{Ly\alpha} = -17$ would count as a significant detection.

This illustrates the type of sophisticated LAH detection predictions that will be possible with \texttt{BlueSi}. In this simple experiment, with a sparse sampling only a single LAH parameter, one can already see the impact varying LAH profiles will have on expected detectability. In the future, we will repeat such tests with grids of multiple varying LAH parameters, such as the halo and compact scale lengths, to derive a robust predicted selection function for given BlueMUSE observations. With more detailed sampling at the limiting-flux end, we will also be able to better probe the relevant noise properties of BlueMUSE.

This will enable the prediction of expected LAE number counts in a given redshift range and observation scheme, with an assumed distribution of intrinsic LAE luminosities. It will also establish predictions for expected lower limits in luminosity detection at a given level of completeness. These two factors will both be informative in designing LAE-targeting surveys with BlueMUSE.

\section{Outlook and work in progress}
\subsection{Rectangular pixels for BlueMUSE}
Unlike MUSE, which samples the field of view in a nearly quadratic grid, the
BlueMUSE instrument will employ rectangular sampling to reduce the number of
optical elements and increase the S/N thanks to a coarser sampling better suited to the PSF size. Simulations are
necessary to study the effects of the sampling on recovering the intrinsic
spatial shapes of objects.

A first test was carried out with a stellar scene that was simulated with
0\farcs55 FWHM in 2D with \texttt{BlueSi} and a fine sampling (50\,mas pixels). We
converted these to BlueMUSE-like pixel tables with $0\farcs2\times0\farcs2$ and
$0\farcs2\times0\farcs3$ pixels, and re-sampled to an output image with
$0\farcs2\times0\farcs2$ or $0\farcs25\times0\farcs25$ grid using the MUSE
pipeline\cite{musepipeline}. The stars were extracted with the
AstroPy\cite{Astropy} \texttt{photutils}\cite{photutils_v1.5.0}
implementation of DAOPhot \texttt{find}\cite{Stetson_1987}, measuring the
roundness and FWHM of the output image, after combining two exposures rotated
by 90\degr{} in this way. As the simulation contained noise at a low level, the
measurements show a certain spread around the expected values. Both roundness
and FWHM were equal within the noise when comparing between quadratic and
rectangular sampling of the observed stellar field.
This test experiment exemplifies the mutual benefits from developing the data reduction pipeline and 
science cube simulation software in parallel.

A more straight forward method that is easier to handle was then built into \texttt{BlueSi}.
Elongated pixels essentially mean a reduced sample rate in one direction per exposure. Since the 
final data cube will be sampled into a grid of spatially square pixels eventually, this means that neighboring pixels
in the direction of the elongation will be stronger correlated to each other.
In the extreme case of a ratio of 2:1, a hypothetical detector pixel could contribute
to two adjacent pixels in the data cube, leading to a correlation of 50\%.
\texttt{BlueSi} incorporates that effect by convolving the Moffat PSF with the appropriate matrix to inflict
the expected correlation. Effectively it reduces the sampling with independent pixels and thereby the spatial resolution 
of the instrument in the direction of elongation.
The impact of an applied rotation pattern for consecutive exposures could then be simulated by creating and co-adding 
two separate cubes, or by applying the corresponding convolution matrix in one go.

In principle, elongated pixels, and therefore a change in resolution in one spatial direction will not affect the shape of the PSF.
However, if the size of the pixel on the sky approaches the FWHM of the PSF you would expect 
an effect because of the flux of the core of the PSF being distributed to a larger area on the sky within the elongated pixel
than the actual incoming PSF. Again, in the extreme case of a FWHM of PSF that is well below the spatial pixel size, the resulting 
shape would directly reflect the pixel aspect ratio and result in elliptical images of point sources.

\begin{figure}[h!]
\begin{center}
\includegraphics[width=0.4\textwidth]{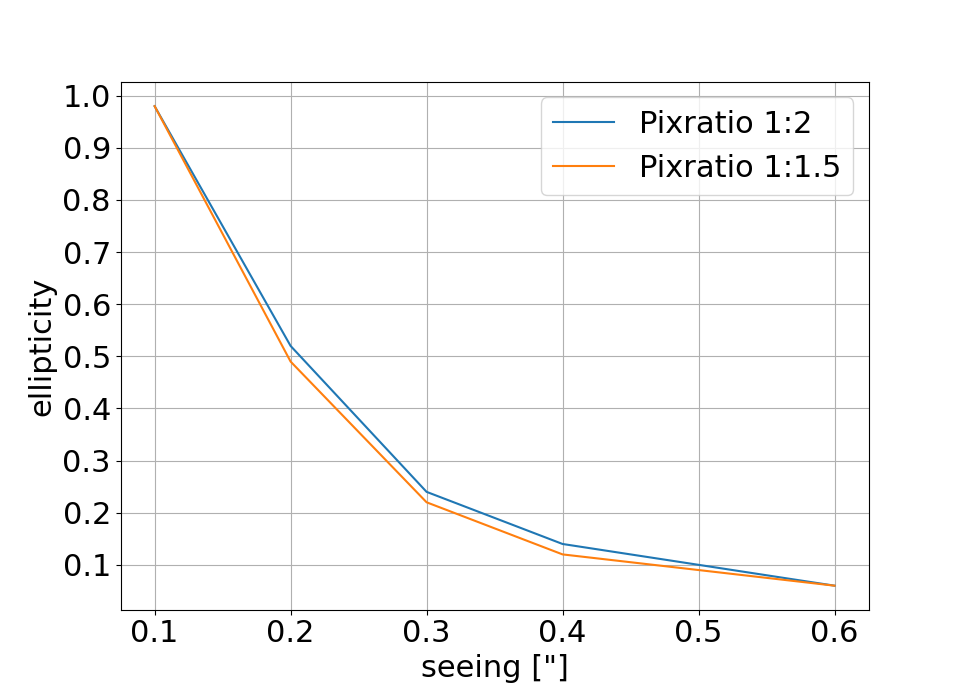}
\end{center}
\caption{The fast decrease in apparent ellipticity of the PSF with seeing.}
\label{fig_ellipse}
\end{figure}

Noise-free images containing several point sources, with different distances
between them were simulated with \texttt{BlueSi} at $0\farcs1<$FWHM$<1\farcs0$. The
profile of an isolated source was then measured with IRAF
\texttt{imexamine}\cite{IRAF_ASCL} which determines FWHM, position, position
angle, and ellipticity, among other parameters. The results for the ellipticity
are most relevant to the effects of rectangular spaxels, and are shown in
Fig.~\ref{fig_ellipse}. It is apparent, that only at unrealistically good
seeing of $<0\farcs3$ the image of the point source is significantly elongated,
at FWHM$\gtrsim 0\farcs6$ that can be expected at blue wavelengths on Paranal,
ellipticities are typically smaller than 0.05. In an exposure with less than
infinite S/N images of stars will therefore appear round, even with rectangular
sampling of the field of view.

With the BlueMUSE specification of $0\farcs2 \times 0\farcs3$ spatial sampling, the expected PSF in the wavelength range of
BlueMUSE will always be at least Nyquist sampled (see \cite{spie_2024richard}) and even for individual exposures without dithering strategies, no
significant deviations from radially symmetric shapes are expected.
\subsection{Effect of covariances in the data}
\label{sec_cov}
\begin{figure}[h!]
\begin{center}
\includegraphics[width=0.3\textwidth]{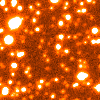}
\includegraphics[width=0.3\textwidth]{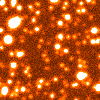}
\includegraphics[width=0.3\textwidth]{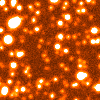}
\end{center}
\caption{A zoom on the NGC 3201 scene at 8221\AA. The three slices show the original VLT/MUSE observation ({\it left}),  the initial \texttt{BlueSi} simulation ({\it center}), and the \texttt{BlueSi} simulation including added correlation with neighboring data-points (spatially and spectrally)
to simulate the covariances ({\it right}). As a consequence, the new simulations are slightly less `grainy' and closer to the noise properties
of the actual reduced data.}
\label{fig_cov}
\end{figure}

\begin{figure}[h!]
\begin{center}
\includegraphics[width=0.3\textwidth]{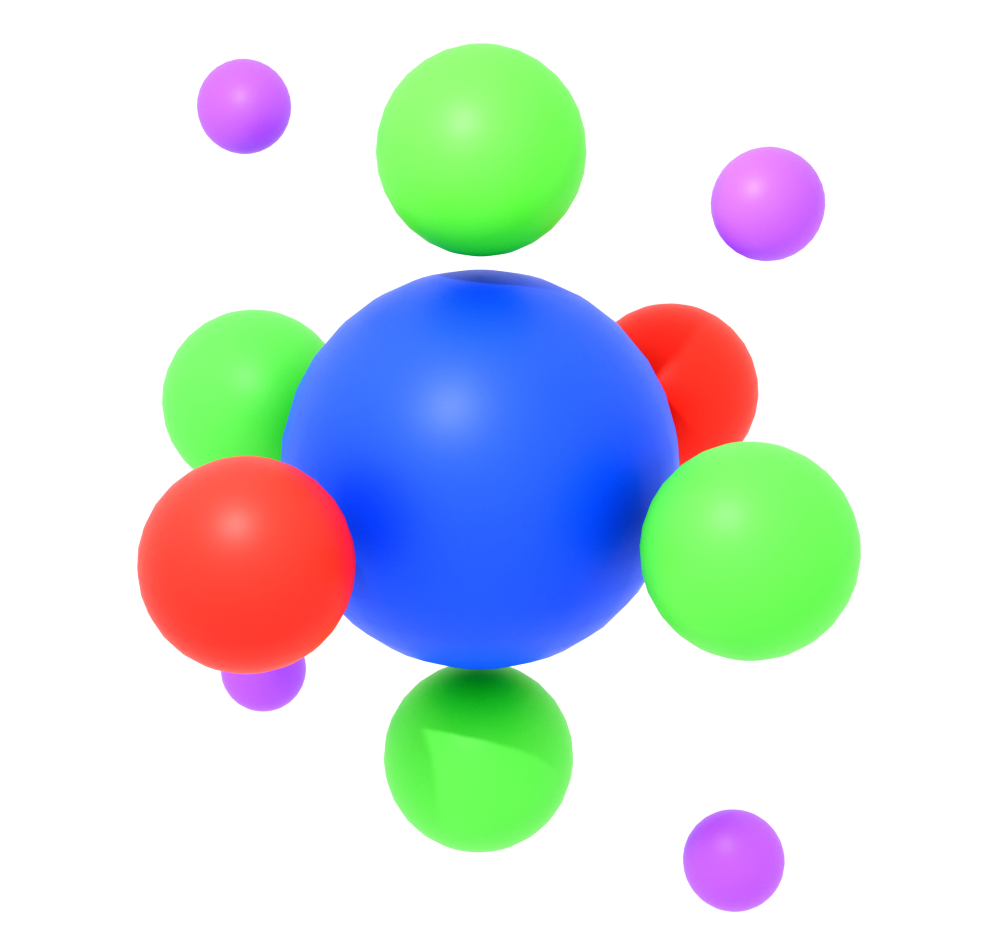}
\end{center}
\caption{The fractional correlation of a central spaxel ({\it blue}) with its direct neighbors scaled to the volumes of the shown spheres in spatial dimension ({\it green}), as well as spectrally ({\it red }).
The influence on the diagonal elements is only depicted for the spatial dimension here ({\it purple}).}
\label{fig_acf_spheres}
\end{figure}

The simulation of the science object as well as the noise are generated 
at the level of the output datacube. The pixel noise is therefore fully independent of its 
neighbors in the data. This differs from the actual data where multiple 
entries in the pix-table can contribute to a certain cube spaxel or vice 
versa, when a single observed flux possibly contributes in part to several data 
cube entries. This is particulary true in the case of rectangular spaxels sampled over a regular grid of square spaxels in the output.
While this is being analyzed in more detail on the level 
of the data reduction pipeline\cite{musepipeline}, \texttt{BlueSi} mimics the 
covariance effects as a result of CCD to regular cube grid resampling as 
shown in the example in Figure \ref{fig_cov}.
The realistic amount of correlation was derived from the average autocorrelation of independent noise on CCD level
which was propagated into a typical data cube of three rotated exposures through the original MUSE data reduction pipeline
(see Fig. \ref{fig_acf_spheres}).

\subsection{Future aspects}
The different sets of simulation runs are currently being analyzed.
While the quality of the simulations is already quite high, there is room for adjustments and improvements on the existing 
implementations, such as the recent additions mentioned in section \ref{sec_cov}.
Furthermore, the content and complexity of the science scenes will be extended, where applicable.
For example the simulation of diffuse gas in \texttt{BlueSi} requires a flexible description of the parameters involved and in
itself represents an important aspect of an early simulation software as a new demand for high resolution synthetic data 
for the extended wavelength range of BlueMUSE develops.
\texttt{BlueSi} separates the creation of the data cube itself and the application of all actual instrument and observation related simulation aspects.
This facilitates the option to use complex science simulations directly as input data, aside from formerly created data within \texttt{BlueSi} itself.
We currently investigate a set of published galaxy simulations to incorporate a new type of science object based on their output models.

One aspect that concerns the code base itself are further optimizations with regard to memory efficiency and parallelization, albeit a full-fledged
simulation is already now completed within minutes to hours.
In close collaboration with the BlueMUSE consortium, additional science scenes are suggested and explored to maintain usability for all covered
science cases. As the technical specifications of BlueMUSE become more and more finalized, \texttt{BlueSi} also aims to simulate more 
subtle effects like a finite precision in the wavelength calibration, variations of the PSF across the FOV as well as a wavelength dependent LSF or
the impact of atmospheric dispersion.
\appendix    
\section{Simulation configuration files}
\subsection{scene config}
\label{sec_scene_config}
\begin{verbatim}
scene_type = 'GC'                     #defines the input handler
objectfile = 'ngc3201_pointing01.csv' #ASCII list specific for scene type
objectdir  = 'ngc3201_pointing01'     #directory of spectra
fov        = 'all'                    #reg/file.reg' or 'all'
maxcount   = -1                       #limit number of objects, -1 = inf
vrad       = yes                      #apply radial shifts from table
vac2air    = yes                      #apply air refraction to model spectra
extinction = yes                      #apply extinction to whole frame
Ebv        = 0.19                     #E(B-V)
simulated_object        = 'NGC 3201_pointing01'
muse_ob_model_spectra   = '2013A&A...553A...6H'
muse_ob_analysis        = '2016A&A...588A.148H'
muse_ob_ngc3201         = '2019A&A...632A...3G'
model_spectra_library   = 'PHOENIX'
model_spectra_packaging = 'Martens,Sven'
scene_file_creator      = 'Wendt,Martin'
\end{verbatim}
\subsection{simulation config}
\label{sec_simulation_config}
\begin{verbatim}
wav_min    = 3500.0              #lower wavelength limit (AA)
wav_max    = 6000.0              #upper wavelength limit (AA)
R          = 4000.0              #Resolution at 5000 AA
d_lambda   = 0.66                #constant bin size (AA)
pixsize    = 0.2                 #detector pixel size (arcsec)
namebase   = 'out/ngc3201_bm'    #name base of generated cubes
instrument = 'BlueMUSE'          #MUSE or BlueMUSE
skytable   = 'skytable_gc.fits'  #sky model
seeing     = 0.63                #parameter for PSF (arsec)
airmass    = 1.1                 #parameter for PSF
psf_beta   = 2.5                 #parameter for PSF
psf_l0     = 22                  #parameter for PSF (m)
raw_dump   = no                  #write cube before PSF/LSF/NOISE
var_ext    = yes                 #add variance extension
do_lsf     = yes                 #render and write LSF 
do_psf     = yes                 #render and write PSF
do_noise   = yes                 #render and write final noise cube 
add_sky    = yes                 #keep sky emission in data?
nexp       = 3                   #number of exposures
texp       = 200                 #exposures time per exposure (sec)
readout    = 3                   #readout noise level (e-)
dark       = 3.0                 #dark current (e-/hour)
saturation = 65535               #16bit saturation
input_type = 'scene/p01.conf'    #scene/file or raw/file BlueSi cube
\end{verbatim}
\acknowledgments 
PMW, NC, and SM gratefully acknowledge support by the BMBF from the ErUM
program (project VLT-BlueMUSE, grants 05A23BAC and 05A23MGA). 
NC also acknowledges funding from the Deutsche Forschungs-gemeinschaft (DFG) - CA 2551/1-1,2.
\bibliography{mainref} 
\bibliographystyle{spiebib} 

\end{document}